\documentclass[aps,twocolumn,superscriptaddress,amsmath,amssymb]{revtex4}

\usepackage{dcolumn}  
\usepackage{bm}       
\usepackage{graphicx} 

\DeclareGraphicsExtensions{.jpg,.pdf,.mps,.png,.eps,.ps,.EPS}

\begin{document}

\title{On the uniform generation of random graphs with prescribed degree sequences}

\author{R. Milo}
\affiliation{Department of Physics of Complex Systems, Weizmann Institute of
Science, Rehovot, Israel 76100}
\affiliation{Department of Molecular Cell Biology, Weizmann Institute of
Science, Rehovot, Israel 76100}

\author{N. Kashtan}
\affiliation{Department of Molecular Cell Biology, Weizmann Institute of
Science, Rehovot, Israel 76100}
\affiliation{Department of Computer Science and Applied Mathematics,
Weizmann Institute of Science, Rehovot, Israel 76100}

\author{S. Itzkovitz}
\affiliation{Department of Physics of Complex Systems, Weizmann Institute of
Science, Rehovot, Israel 76100}
\affiliation{Department of Molecular Cell Biology, Weizmann Institute of
Science, Rehovot, Israel 76100}

\author{M. E. J. Newman}
\affiliation{Department of Physics and Center for the Study of Complex
Systems, University of Michigan, Ann Arbor, MI 48109--1120, U.S.A.}

\author{U. Alon}
\affiliation{Department of Physics of Complex Systems, Weizmann Institute of
Science, Rehovot, Israel 76100}
\affiliation{Department of Molecular Cell Biology, Weizmann Institute of
Science, Rehovot, Israel 76100}

\begin{abstract}
Random graphs with prescribed degree sequences have been widely
used as a model of complex networks.  Comparing an observed
network to an ensemble of such graphs allows one to detect
deviations from randomness in network properties.  Here we briefly
review two existing methods for the generation of random graphs
with arbitrary degree sequences, which we call the ``switching''
and ``matching'' methods, and present a new method based on the
``go with the winners'' Monte Carlo method.  The matching method
may suffer from nonuniform sampling, while the switching method
has no general theoretical bound on its mixing time.  The ``go
with the winners'' method has neither of these drawbacks, but is
slow. It can however be used to evaluate the reliability of the
other two methods and, by doing this, we demonstrate that the
deviations of the switching and matching algorithms under
realistic conditions are small compared to the ``go with the
winners'' algorithm. Because of its combination of speed and
accuracy we recommend the use of the switching method for most
calculations.
\end{abstract}

\pacs{05, 89.75}
\maketitle

\section{Introduction}
In the rapidly growing literature on the modeling of complex
networks one of the most important classes of network models is
the random graph~\cite{Bollobas2001}.  One well-studied such model
is the model consisting of the ensemble of all graphs that have a
given degree sequence~\cite{Bender,Molloy 1995,Molloy 1998,Newman
2001,Chung_diameter}, and this model has proved useful in
understanding a variety of network properties.  Realistic
applications often require that we restrict ourselves to graphs
with no multiple edges between any vertex pair and no self-edges.
Unfortunately, both the analytic and numerical study of such
networks is known to present
challenges~\cite{Bender,Snijders,Rao,Roberts,Kannan,chen,Itzkovitz,MSZ02,Park}.
In this short paper we consider computer algorithms for generating
graphs uniformly from this ensemble.  We are concerned primarily
with directed graphs, since the examples we will consider are
directed, but the concepts discussed generalize in a
straightforward fashion to the undirected case also.

There are two algorithms in common use for the generation of
random graphs with single edges.  We will refer to them as the
\textit{switching algorithm}~\cite{Rao,Roberts,Newman 2002,Maslov
2002,Stone,Shen or 2002,Milo 2002} and the \textit{matching
algorithm}~\cite{Molloy 1995,Newman 2001,Milo 2002}.  We argue
that, under certain circumstances, both of these algorithms can
generate a nonuniform sample of possible graphs.  We then present
a new algorithm based on the Monte Carlo procedure known as
\textit{go with the winners}~\cite{Aldous,Grassberger}, which
generates uniformly sampled graphs.  We compare the three methods
in the context of a particular network problem---estimation of the
density of commonly occurring subgraphs or \textit{motifs}---and
show that, in this context, the difference between them is small.
This result is of some practical importance, since the ``go with
the winners'' algorithm, although statistically correct, is slow,
while the other two algorithms are substantially faster.

\begin{table*}
\setlength{\tabcolsep}{2.5pt}
\begin{tabular}{l|c|c|c|c|c|c|c|c|c|c|c|c}
Network & \multicolumn{3}{c|}{E. coli transcription} &
 \multicolumn{3}{c|}{yeast transcription} &
 \multicolumn{3}{c|}{C. elegans neurons}  &
 \multicolumn{3}{c}{electronic circuit S15850 \cite{Brglz}} \\
\cline{2-13}
 & mean & s.d. & $Z$ & mean & s.d. & $Z$ & mean& s.d. & $Z$ & mean & s.d. & $Z$ \\
\hline
``go with the winners'' & 7.57(5) & 3.05(3) & 10.6(1) & 11.06(6) & 3.60(4) & 14.1(2) & 88(1) & 10.7(7) & 3.4(3) & 2.20(5) & 1.48(3) & 284(6) \\
switching & 7.63(9) & 3.05(6) & 10.5(2) & 11.0(1) & 3.71(7) & 13.7(3) & 88.3(3) & 10.1(2) & 3.6(1) & 2.24(5) & 1.47(3) & 286(6) \\
matching & 7.67(9) & 2.98(6) & 10.8(2) & 11.1(1) & 3.67(7) & 13.8(3) & 94.5(3) & 10.0(2) & 3.0(1) & 2.21(5) & 1.45(3) & 290(6) \\
\end{tabular}
\caption{Mean and standard deviation (s.d.)\ of the number of
appearances of the feed-forward loop subgraph in random networks
with degree sequences the same as the real world networks studied
in~\cite{Milo 2002}.  We used between $1000$ and $10\,000$ random
networks for each measurement. $Z$-scores are the number of
standard deviations by which the real network deviates from the
average of the random ensemble. Standard errors are shown in
parentheses.} \label{Table1}
\end{table*}

\begin{figure}[!hbp]
\begin{center}
\resizebox{7cm}{!}{\includegraphics{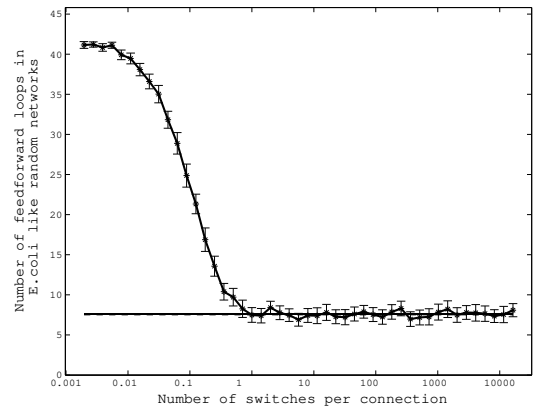}}
\end{center}
\caption{Starting with the transcription network of E. coli, the
network is randomized using the switching algorithm described in
the text.  We plot the number of feed-forward loops in the
randomized networks vs.\ number of switches performed per edge in
the graph.  The dashed line is the expected asymtotic value
obtained using the ``go with the winners'' algorithm.  Each point
is an average over one hundred repetitions of the calculation.
Error bars are $\pm3$ standard deviations.  The randomized network
reaches the equilibrium value around one switch per edge on
average.  Similar results are obtained for other networks and
other motifs.} \label{fig2}
\end{figure}

\section{Algorithms}
In this section we describe the three algorithms under consideration.

\subsection{Switching algorithm}
First, we describe the switching algorithm, which uses a Markov chain to
generate a random graph with a given degree sequence
~\cite{Rao,Roberts,Newman 2002,Maslov 2002,Stone,Shen or 2002,Milo
2002}. For simplicity, we discuss directed networks with no mutual
edges (vertex pairs with edges running in both directions between
them).  The case with mutual edges is a simple
generalization~\cite{Roberts}.

The method starts from a given network and involves carrying out a
series of Monte Carlo switching steps whereby a pair of edges
$(A\to B,C\to D)$ is selected at random and the ends are exchanged
to give $(A\to D,C\to B)$. However, the exchange is only performed
if it generates no multiple edges or self-edges; otherwise it is
not performed.  The entire process is repeated some number $QE$
times, where $E$ is the number of edges in the graph and $Q$ is
chosen large enough that the Markov chain shows good mixing.
(Exchanges that are not performed because they would generate
multiple or self-edges are still counted to insure detailed
balance~\cite{footnote}.)

This algorithm works well but, as with many Markov chain methods,
suffers because in general we have no measure of how long we need
to wait for it to mix properly.  Theoretical bounds on the mixing
time exist only for specific near-regular degree
sequences~\cite{Kannan}. We empirically find, however, that for
many networks, values of around $Q=100$ appear to be more than
adequate (see Fig.~\ref{fig2}).

\subsection{Matching algorithm}
An alternative approach is the matching algorithm~\cite{Molloy
1995,Newman 2001,Milo 2002}, in which each vertex is assigned a
set of ``stubs'' or ``spokes''---the sawn-off ends of incoming and
outgoing edges---according to the desired degree sequence.  (One
can also assign mutual-edge stubs for networks that include such
edges.)  Then in-stubs and out-stubs are picked randomly in pairs
and joined up to create the network edges.  If a multiple or
self-edge is created, the entire network is discarded and the
process starts over from scratch.

This process will correctly generate random directed graphs with
the desired properties.  Unfortunately, however, many real-world
networks have a heavy-tailed degree distribution that includes a
small minority of vertices with high degree.  All other things
being equal, the expected number of edges between two such
vertices will often exceed one, making it unlikely that the
procedure above will run to completion, except in the rarest of
cases.  To obviate this problem a modification of the method can
be used in which, following selection of a stub pair that creates
a multiple edge, the network is not discarded, and an alternative
stub pair is selected at random.  In general this method generates
a biased sample of possible networks~\cite{King} but, as we will
show, not significantly so for our purposes (see
Table~\ref{Table1}).

\begin{figure}
\begin{center}
\resizebox{\columnwidth}{!}{\includegraphics{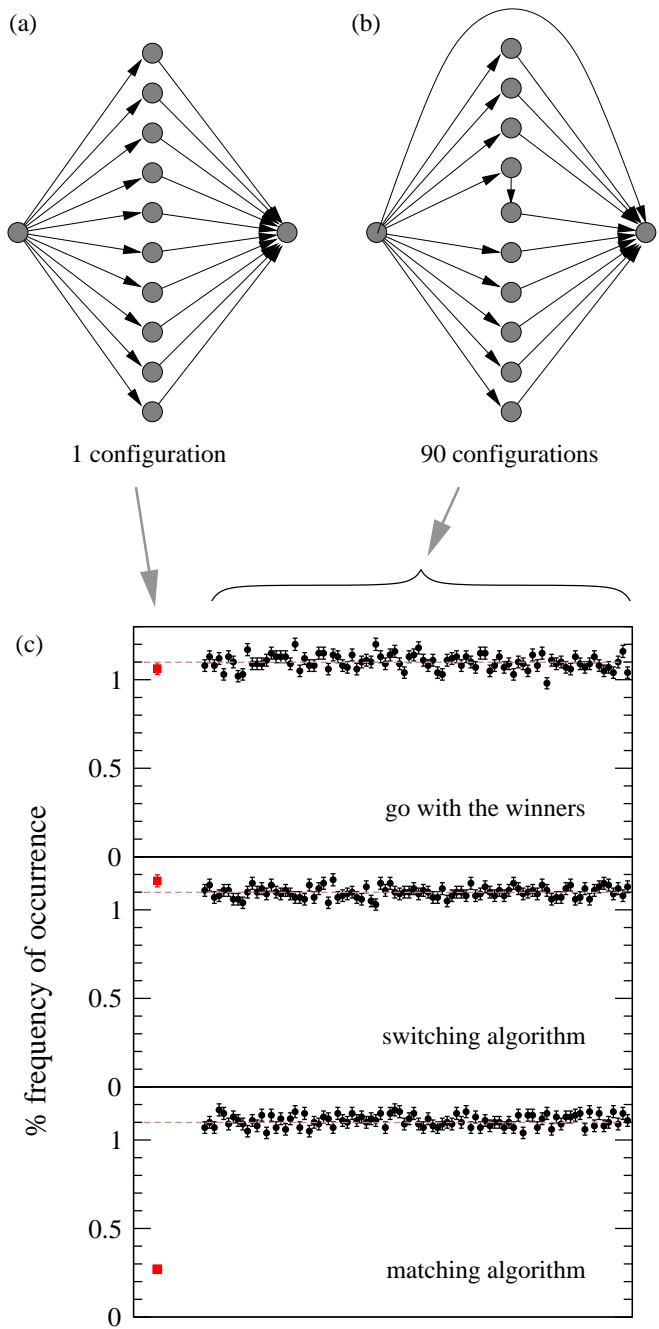}}
\end{center}
\caption{Uniformity tests of the three algorithms on a toy
network.  Panels~(a) and~(b) depict the two types of topologies of
the 91 random networks studied, one of them like~(a) and 90
like~(b).  Panel~(c) shows the frequency with which each
configuration is sampled by our three algorithms.  $100\,000$
graphs were generated with each algorithm, and the figure shows
the fraction of graphs of each type generated.  If sampling were
uniform, each should appear with probability $\frac{1}{91}$, which
is indicated by the dotted lines.  The ``go with the winners'' and
switching algorithms sample uniformly within sampling error,
passing both the Kolmogorov--Smirnoff and Lillie Gaussian tests.
The matching algorithm under-samples the unique
configuration~(a).} \label{hub1}
\end{figure}

\subsection{Go with the Winners algorithm}
The ``go with the winners'' algorithm is a non-Markov-chain Monte
Carlo method for sampling uniformly from a given
distribution~\cite{Aldous,Grassberger}. When applied to the
problem of graph generation, the method is as follows. We consider
a colony of $M$ graphs.  As with the matching algorithm, we start
with the appropriate number of in-stubs and out-stubs for each
vertex and repeatedly choose at random one in-stub and one
out-stub from the graph and link them together to create an edge.
If a multiple edge or self-edge is generated, the network
containing it is removed from the colony and discarded.  To
compensate for the resulting slow decline in the size of the
colony, its size is periodically doubled by cloning each of the
surviving graphs; this cloning step is carried out at a
predetermined rate chosen to keep the size of the colony roughly
constant on average.  The process is repeated until all stubs have
been linked, then one network is chosen at random from the colony
and assigned a weight:
\begin{eqnarray}
W_i=2^{-c}\frac{m}{M},
\label{eq2}
\end{eqnarray}
where $c$ is the number of cloning steps made and $m$ is the number of
surviving networks.  The mean of any quantity $X$ (for example, the number
of occurrences of a given subgraph) over a set of such networks is then
given by
\begin{eqnarray}
\frac{\sum_{i}{{W_i}{X_i}}}{{\sum_{i}{W_i}}},
\label{eq3}
\end{eqnarray}
where $X_i$ is the value of $X$ in network~$i$.

\section{Comparison of algorithms}
In Fig.~\ref{hub1} we show a comparison of the performance of our three
algorithms when applied to a simple toy network.  The network consists of
an out-hub with ten outgoing edges, an in-hub with ten incoming edges, and
ten nodes with one incoming edge and one outgoing edge each.  Given this
degree sequence, there are just two distinct network topologies with no
multiple edges, as shown in Fig.~\ref{hub1}a and~\ref{hub1}b.  There is
only a single way to form the network in \ref{hub1}a, but there are 90
different ways to form~\ref{hub1}b.

We generated $100\,000$ random networks using each of the 3
methods described here and the results are summarized in
Fig.~\ref{hub1}c.  As the figure shows, the matching algorithm
introduces a bias, undersampling the configuration of
Fig.~\ref{hub1}a.  This is a result of the dynamics of the
algorithm, which favors the creation of edges between hubs.  The
switching and ``go with the winners'' algorithms on the other hand
sample the configurations uniformly, generating each graph an
equal number of times within the measurement error on our
calculations.  The ``go with the winners'' algorithm truly samples
the ensemble uniformly but is far less efficient than the two
other methods.  The results given here indicate that the switching
algorithm produces essentially identical results while being a
good deal faster.  The matching algorithm is faster still but
samples in a measurably biased way.

Now consider the study of network motifs.  We are interested in
knowing when particular subgraphs or motifs appear significantly
more or less often in a real-world network than would be expected
on the basis of chance, and we can answer this question by
comparing motif counts to random graphs. Some results for the case
of the ``feed-forward loop'' motif~\cite{Shen or 2002,Milo 2002}
are given in Table~\ref{Table1}.  In this case the densities of
motifs in the real-world networks are many standard deviations
away from random, which suggests that any of the present
algorithms is adequate for generating suitable random graphs to
act as a null model, although the ``go with the winners'' and
switching algorithms, while slower, are clearly more satisfactory
theoretically.  The matching algorithm was measurably nonuniform
for our toy example above, but seems to give better results on the
real-world problem.

Overall, our results appear to argue in favor of using the
switching method, with the ``go with the winners'' method finding
limited use as a check on the accuracy of sampling.  Accuracy
checks are also supplied by analytical estimates for subgraph
numbers~\cite{Itzkovitz}. Numerical results in~\cite{Milo
2002,Milo 2004,Itzkovitz} were done using the switching algorithm.

\section{Conclusions}
In this paper we have compared three algorithms for generating
random graphs with prescribed degree sequences and no multiple
edges or self-edges.  Two of the three have been used previously,
but suffer from nonuniformity in their sampling properties, while
the third, a method based on the ``go with the winners'' Monte
Carlo procedure, is new and samples uniformly but is quite slow.
Of the two older algorithms, we show that one, which we call the
``matching'' algorithm, has measurable deviations from uniformity
when compared to the ``go with the winners'' method, although for
graphs typical of practical studies these deviations are small
enough to make no significant difference to most previously
published results.  The other older algorithm, which we call the
``switching'' algorithm and which is based on a Markov chain Monte
Carlo method, samples correctly in the limit of long times and in
practice is found to give good results when compared with the ``go
with the winners'' method.  Overall, therefore, we conclude that
the switching algorithm is probably the algorithm of choice, with
the ``go with the winners'' algorithm finding a supporting role as
a check on uniformity, although its slowness
makes it impractical for large-scale use.\\
We thank Oliver D. King for discussions and for pointing out and
demonstrating that the matching algorithm of the supplementary
online material of \cite{Milo 2002} does not uniformly generate
simple graphs.


\begin{thebibliography}{99}
\bibitem{footnote}
There are singular networks where the Markov process is not
ergodic.  This lack of ergodicity can be removed by making small
modifications as in~\cite{Rao}, and by choosing the number of
switching steps per edge to be randomly distributed around Q.
\end{thebibliography}
\end{document}